\documentclass[twocolumn,showpacs,preprintnumbers,amsmath,amssymb]{revtex4}
\usepackage{graphicx}
\usepackage{dcolumn}
\usepackage{bm}

\newcommand{\ba}{\begin{eqnarray}}
\newcommand{\ea}{\end{eqnarray}}
\newcommand{\bse}{\begin{subequations}}
\newcommand{\ese}{\end{subequations}}
\newcommand{\dd}{\hbox{d}}

\newcommand{\M}{{\cal {M}}}
\newcommand{\CH}{{\cal {H}}}
\newcommand{\F}{{\cal {F}}}

\newcommand{\CZ}{{\cal {Z}}}
\newcommand{\la}{\langle}
\newcommand{\ra}{\rangle}

\begin{document}

\title{Inhomogeneous models of interacting dark matter and
dark energy.}
\author{Roberto A. Sussman$^\dagger$, Israel Quiros$^\ddagger$ and Osmel Mart\'\i n
Gonz\'alez$^\ddagger$}

\address
{$^\dagger$ Instituto de Ciencias Nucleares,  Apartado Postal
70543, UNAM, M\'exico DF, 04510,
M\'exico.\\$^\ddagger$Departamento de F\'\i sica,  Universidad
Central de las Villas, Santa Clara, Cuba.}

\email{sussman@nuclecu.unam.mx;israel@uclv.edu.cu;osmel@uclv.edu.cu}

\begin{abstract}
We derive and analyze a class of spherically symmetric cosmological models whose
source is an interactive mixture of inhomogeneous cold dark matter (DM)  and a
generic homogeneous dark energy (DE) fluid. If the DE fluid corresponds to a quintessense
scalar field, the interaction term can be associated with a well motivated non--minimal
coupling to the DM component. By constructing a suitable volume average of the
DM component we obtain a Friedman evolution equation relating this
average density with an average Hubble scalar, with the DE component playing
the role of a repulsive and time-dependent $\Lambda$ term.  Once we select an 
``equation of state'' linking the energy density ($\mu$) and pressure ($p$) of
the DE fluid, as well as a free function governing the radial dependence, the models
become fully determinate and can be applied to known specific DE sources, such as
quintessense scalar fields or tachyonic fluids.  Considering the simple equation of state
$p=(\gamma-1)\,\mu$ with $0\leq\gamma <2/3$, we show that the free parameters and
boundary conditions can be selected for an adequate description of a local DM overdensity
evolving in a suitable cosmic background that accurately fits current observational data.
While a DE dominated scenario emerges in the asymptotic future, with total $\Omega$ and
$q$ tending respectively to 1 and -1/2 for all cosmic observers, the effects of
inhomogeneity and anisotropy yield different local behavior and evolution rates for these
parameters in the local overdense region.  We suggest that the models presented can be
directly applied to explore the effects of various DE formalisms on local DM cosmological
inhomogeneities.
\end{abstract}

\maketitle

\section{Introduction}
Observational data on Type Ia supernovae strongly suggests that the universe
is expanding at an accelerated rate\cite{newuniv1,newuniv2}. This effect has lead to
the widespread assumption that the inventary of cosmic matter--energy could
contain, besides baryons, photons, neutrinos and cold dark matter (DM)\footnote{We shall
assume henceforth that DM is of the ``cold'' variety, {\it i.e.} CDM.}, an
extra contribution generically known as ``dark energy'' (DE), whose kinematic effect
could be equivalent to that of a fluid with negative pressure. While the
large scale dynamics of the main cosmic sources (DE and DM) is more or less
understood, their fundamental physical nature is still a matter for debate, thus
various physical explanations have been suggested. Cold DM is usually
conceived as a collisionless gas of supersymmetric particles
(neutralinos), while DE can be modelled as a ``cosmological constant'', quintessense
scalar fields, tachyonic fluids, generalized forces, etc\cite{Padma1, Lima1}.  The
standard approach is mostly to consider a Friedman-Lema\^\i tre-Robertson-Walker
(FLRW) metric, with linear pertubations, making also the simplest assumption that DE only
interacts gravitationally with DM. However, there are still some unresolved issues, such
as the so--called ``coincidence problem'', concerning the odd apparent fact that the
critical densities of DM and DE approximately coincide in our cosmic
era~\cite{amendola,coinc1}. Aiming at a solution to this problem and bearing in mind our
ignorance on the fundamental physics of DM and DE, various models have been proposed
recently which include asorted forms of interaction between these
sources~\cite{Q-int1,Q-int2,Q-int3,Q-int4,Q-int5}.

It is customarily assumed that DE dominates large scale cosmic dynamics, so
that DM inhomogeneities in galactic clusters and superclusters can be considered
a local effect or can be treated by means of linear perturbations in a FLRW
background. Thus, a reasonable generalization of existing models could be to
assume inhomogeneous DM interacting with homogenous DE, so that large
scale dynamics is governed by the latter. We propose in this paper a class of
analytic models which provide a reasonable description of inhomogeneous DM
interacting with a generic homogeneous DE source. The models are based
on the spherically symmetric subcase of the Szafron--Szekeres exact solucions of
Einstein's field equations for a perfect fluid source\cite{kras}. However, the
underlying geometry of the models we present can be easily generalized to include
non--spherical symmetries or even the case without any isometry, since, in
general, Szafron--Szekeres solutions do not admit Killing vectors. 

The prefect
fluid in Szafron--Szekeres solutions in their original conception is
characterized by a non--rotating geodesic 4-velocity field, so that in the
comoving frame matter--energy density is inhomogeneous, while pressure depends
only on cosmic time. As we show in section II, it is straightforward to
re--interpret this fluid as a mixture of an inhomogeneous dust component plus a
homogeneous fluid. Such mixture have been considered
previously~\cite{old2F1,old2F2} but in the context of mixtures of baryons and
radiation. We consider in this paper only the type of models examined
in~\cite{old2F2}, by assuming the homogeneous fluid to describe a generic DE
source, while the dust component corresponds to inhomogeneous DM, all of which
is a reasonable assumption since the dynamical effects of quintessence mostly
become dominant in very large scales, larger than the ``homogeneity scale''
(100--300 Mpc), while DM (galactic clusters and superclusters) is very
inhomogeneous at scales of this magnitude and smaller. By conveniently rescaling
the free parameters and calculating relevant physical and geometric quantities,
we show in sections III, IV and V that the dynamical equation characterizing the
models is analogous to a Friedman equation in which the average DM density 
evolves in the presence of a generic (still undetermined) time--dependent
cosmological constant or ``Lambda field''. By looking at the regularity
conditions, singularities and asymptotic behavior along timelike and spacelike directions,
we show in section VI that the DE and DM mixture components behave as needed for a
reasonable cosmological model complying with observations: for all fundamental
observers the DE source dominates over DM in the asymptotic timelike future, while
boundary conditions determine how the local ratio of DE and DE changes along the rest
frames of the fundamental observers. However, these frames (hypersurfaces of constant
cosmic time) have zero curvature, hence the inhomogeneities in the models are more
suitable for a study of large scale inhomogeneities than local structure formation, since
even overdense regions homogeneize and isotropize towards an asymptotic DE dominated
scenario. The main observational parameters are  appropriately defined and calculated for
an inhomogeneous and anisotropic spacetime  in section VII.

In order to determine the time evolution of the sources, we need to assume a
physical model, or ``equation of state'' for the generic DE source (the
homogeneous fluid). Thus, we assume in section VIII a simple ``gamma law'' equation of
state of the form $p=(\gamma-1)\,\mu $, where $p,\,\mu$ are the pressure and
matter--energy density of the DE source. Such an equation of state leads
to a DE homogeneous fluid evolving like a FLRW fluid with flat spacelike sections
with a scaling law of the form $\mu\propto t^{-2}$, which is compatible with a
scalar field with an exponential potential~\cite{CJSF}. Although this is a very simple
type of DE source, it yields analytic forms for the DM density, observational
parameters and allother  relevant physical and geometric
quantities.

The assumption of a gamma law equation of state fully determines the time dependence of
all relevant quantities, but the free parameters governing spacial dependence are
specified in section IX. Suitable boundary conditions can always be selected allowing
for a description of a local DM overdense region in a DE dominated cosmic background that
accurately complies with observational constraints on observational parameters: $\Omega$
for DM and for DE and the deceleration parameter $q$. We provide in this section a
full graphical illustration of the interplay between ``local'' and ``cosmic background''
effects on these observational parameters:  for example, anisotropy emerges in the local
dependence of these quantities on the  ``off-center observation angle'' $\psi$, while
inhomogeneity leads to local conditions in the overdense region (DM dominates over DE and 
$q$ is positive) that are different  from those of the cosmic background: DE
dominates and $q<0$, as required by an ``accelerated'' universe whose large scale
dynamics is dominated by a repulsive force associated with DE. 

The issue of the interaction between DE and DM is dealt with in section X.
We show that the individual momentum--energy tensors for DM and DE are not independently
conserved, thus the models are incompatible with these components interacting only
gravitationally. However, if we assume the DE fluid to be a scalar field quintessense
type of source, then the models can accomodate various prescriptions for a DE--DM
interaction, like those proposed in the
literature~\cite{Q-int1,Q-int2,Q-int3,Q-int4,Q-int5}. Finally, in section XI we present a
discussion and summary of our results.

\section{Field equations for a class of inhomogeneous cosmologies.}

We start with a spherically symmetric inhomogeneous
spacetime, the spherical subcase of the Szafron--Szekeres solutions~\cite{kras}, which
can be described by the Lemaitre--Tolman--Bondi (LTB) metric element
\begin{equation} ds^2 \ = \
-c^2\,dt^2+\frac{Y'{}^2}{1-K}\,dr^2+Y^2\,\dd\Omega^2,\label{ltbmetric}
\end{equation}
with $K=K(r),\,\,Y=Y(t,r)$, \,\, $\dd \Omega^2 \equiv
\dd\theta^2+\sin^2\theta\,\dd\varphi^2 $ and a prime denotes derivative with
respect to
$r$. As matter source we consider a perfect fluid
\begin{equation}T^{ab} \ = \
(e+p)\,u^a\,u^b+p\,g^{ab},\label{TabPF}\end{equation}
where $e$ and $p$ are the matter--energy density
and total (effective) pressure. In a comoving representation with 4-velocity
$u^a=\delta^a_t$ the field equations for (\ref{ltbmetric}) and (\ref{TabPF}) are
\bse\ba G^r_r = \kappa\,p &=& -\frac{K}{Y^2}-\frac{\dot
Y^2}{Y^2}-\frac{2\ddot Y}{Y},\label{G11}\\
G^\theta_\theta = G^\phi_\phi=\kappa\,p &=& -\frac{K'}{2YY'}-\frac{\dot
Y\dot Y'}{YY'}-\frac{\ddot Y}{Y}-\frac{\ddot Y'}{Y'},\label{G22}\\
-G^t_t = \kappa\,e &=& \frac{K}{Y^2}+\frac{\dot
Y^2}{Y^2}-\frac{2Y'\dot Y'}{YY'}-\frac{K'}{YY'},\label{G44}\ea\ese
where $\kappa=8\pi G/c^4$ and a dot denotes derivative with respect to $ct$.
Since the left hand sides of (\ref{G11}) and (\ref{G22}) are identical, we
obtain the condition $G^\theta_\theta-G^r_r=0$, which yields
\begin{equation}\frac{K}{Y^2}+\frac{\dot
Y^2}{Y^2}+\frac{2\ddot Y}{Y} \ = \ -\kappa\,p(t).\label{CPI}\end{equation}
an expected result since the comoving 4-velocity for (\ref{ltbmetric}) is
a timelike geodesic vector and so is incompatible with spacelike gradients of
$p$. If we keep $p(t)$ arbitrary, then we cannot obtain a simple integral of
(\ref{CPI}), but we can transform this condition into
\begin{equation}\ddot
X+\frac{3}{4}\,\kappa\,p\,X+\frac{3}{4}\,\frac{K}{X^{1/3}} \ = \
0,\label{ddotX}\end{equation}
where $X=Y^{3/2}$, while matter--energy density in (\ref{G44}) becomes
\begin{equation}\kappa\,e \ = \ \frac{4}{3}\,\frac{\dot
X\,\dot
X'}{X\,X'}+\frac{K}{X^{4/3}}+\frac{3}{2}\,\frac{K'}{X^{1/3}\,X'}.
\label{eqe}\end{equation}
Hence, once we choose $p$ and $K$, we can find $X$ by integrating (\ref{ddotX}) and
then $e$ becomes determined by means of (\ref{eqe}). The solutions of Einstein's
field equations associated with (\ref{ltbmetric}), (\ref{ddotX}) and (\ref{eqe})
are the spherically symmetric subcase of ``class 1'' Szekeres--Szafron
solutions~\cite{kras}. However, for $K\ne 0$ there are no analytic
solutions for the nonlinear equation (\ref{ddotX}), while finding a
meaningful equation of state relating $p(t)$ and
$e(t,r)$ is difficult.

\section{A mixture of dark matter and dark energy.}

An alternative approach is to consider only the case $K=0$ (hypersurfaces $t=$ constant
have zero curvature) and to assume that total matter--energy density decomposes as
\begin{equation} e(t,r) \ = \rho(t,r)\,c^2 + \mu(t),\label{edesc}\end{equation}
so that we can re--interpret the momentum--energy tensor (\ref{TabPF}) as the
mixture
\ba T^{ab}&=&T^{ab}_{(1)}+T^{ab}_{(2)},\qquad \textrm{with:}\nonumber\\
T^{ab}_{(1)}&=&\rho\,c^2\,u^a\,u^b,\quad
T^{ab}_{(2)}=(\mu+p)\,u^a\,u^b+p\,g^{ab}.\nonumber\\
\label{Tdesc}\ea
So, it is now more natural to choose $\mu$ and $p$, as matter--energy
density and total pressure of a homogeneous fluid describing a specific physical
system. Then, integrating the linear equation
\begin{equation}\ddot
X+\frac{3}{4}\,\kappa\,p\,X=0,\label{ddotX0}\end{equation}
we can determine the metric and the rest--mass density of the inhomogeneous dust
component
\begin{equation}\kappa\,\rho\,c^2 = \frac{\frac{4}{3}\,\dot
X\dot X'-\kappa\,\mu\,XX'}{XX'}.\label{eqrho}\end{equation}
This approach has been used in the past for modeling mixtures of baryons and
radiation~\cite{old2F1,old2F2}, but it can be equally useful to study the interaction
between cold DM (the inhomogeneous dust: $T^{ab}_{(1)}$) and a generic (yet unspecified)
type of DE (the homogeneous fluid $T^{ab}_{(2)}$ with negative pressure). 
Because of their construction, the two energy--momentum tensors are not separately
conserved: \, $T^{ab}_{(1)}{}_{;b}=-T^{ab}_{(2)}{}_{;b}\ne 0$, hence we must have a
non--minimal coupling between DM and DE (see section X).\\

\section{Determination of free parameters}

Since (\ref{ddotX0}) is a second order lineal partial differential equation, its
solutions have the form
\begin{equation} X \ = \ V(t)\,\alpha(r)+W(t)\,\beta(r),\label{Xform}\end{equation}
where $\alpha(r),\,\beta(r)$ are the two ``integration constants'' that follow from the
integration of (\ref{ddotX0}) and the functions
$V,\,W$ satisfy: $\ddot V/V=\ddot W/W =(3/4)\,\kappa\,p$. Since we can
arbitrarily relabel the radial coordinate $r$, no loss of generality is involved
if we relabel these functions as
\bse\label{defs}\ba V=R^{2/3},\qquad W=T\,R^{2/3},\\
\qquad \alpha=r^{2/3},\qquad \beta = r^{2/3}\,f \ea\ese
where $R(t),\,T(t)$ and $f(r)$ are arbitrary functions. Inserting (\ref{Xform})
and (\ref{defs}) in (\ref{ddotX0}) and demanding consistency with
(\ref{CPI})--(\ref{eqe}), we obtain the equations that (given a choice of
$p=p(\mu)$) determine
$R$ and $T$
\ba \kappa\,p \ &=& \ -\frac{\dot R^2}{R^2}-\frac{2\ddot R}{R},\label{eq_p}\\
\kappa\,\mu \ &=& \ \frac{3 \dot
R^2}{R^2},\label{eq_mu}\\
\dot T \ &=& \ c_0\,H_0\frac{R_0^3}{R^3},\label{eq_eps}\ea
where $c_0$ is a dimensionless constant, $R_0=R(t_0)$ and $H_0$ can be identified with the
Hubble timescale at $t_0$. The form of equations (\ref{eq_p}) and (\ref{eq_mu}) is
identical to the field equations of a FLRW spacetime with flat space sections, this
suggest that we identify $\mu$ and $p$ with variables somehow associated with a FLRW
background. Once we choose an ``equation of state'' $p=p(\mu)$ corresponding to a
specific DE model for the homogeneous fluid (for example, a scalar field), we can find
$R$ by integrating (\ref{eq_p}) and (\ref{eq_mu}), and then $T$ by
integrating (\ref{eq_eps}). 

An alternative approach follow by defining what would be a ``Hubble factor'' for
an homogeneous fluid: 
\begin{equation}H \equiv  \frac{\dot R}{R},\label{Hhat}\end{equation}
allowing us to combine
equations (\ref{eq_p}) and (\ref{eq_mu}) into
\begin{equation}\dot H \ = \ -\frac{\kappa}{2}\,(\mu+p),\qquad
\kappa\,\mu \ = \ 3\,H^2,\label{Hdot}\end{equation}
so that, once the ``equation of state''
$p=p(\mu)$ is prescribed, we get $p=p(H)$. The functions $H$ and $R$ follow by
integration of (\ref{Hdot}) and (\ref{Hhat}), which is equivalent to
integrating (\ref{eq_p}) and (\ref{eq_mu}).

Once we select $f$ (after solving (\ref{eq_p})--(\ref{eq_mu})  or
(\ref{Hdot})--(\ref{Hhat})), the metric (\ref{ltbmetric}) and dust density (\ref{eqrho}),
as well as any other geometric or physical variables, become fully determined. Under the
parametrization given by (\ref{Xform}) and (\ref{defs}), the metric (\ref{ltbmetric}) and
DM mass--energy density (\ref{eqrho}) take the form
\begin{widetext}
\ba\dd s^2 \ = \ -c^2\,\dd
t^2+R^2\,\left[\,\frac{(1+F\,T)^2}{(1+f\,T)^{2/3}}\,\dd
r^2+r^2\,(1+f\,T)^{4/3}\,
\dd\Omega^2\right],\label{metric}\ea\end{widetext}

\ba\kappa\,\rho\,c^2 \ &=& \
\frac{4\,(3\,T\,H+\dot T)\,f\,F+6\,
H\,(f+F)}{3\,(1+f\,T)\,(1+F\,T)}\,
\dot T\nonumber\\
&=&
\frac{2\,c_0\,H_0\,R_0^3}{3\,R^6}\,\frac{\frac{d}{dt}[\,(2\,T\,f\,F+f+F)\,R^3\,]}
{(1+f\,T)\,(1+F\,T)}\label{eq_rho}\ea
where we used (\ref{eq_eps}) and
\begin{equation}
F  \equiv  f+\frac{2}{3}\,r\,f\,'.\label{Fdef}\end{equation}
Other important quantities are the expansion kinematic scalar
$\Theta=u^a\,_{;a}$ and the traceles symmetric shear tensor
$\sigma_{ab}=u_{[a;b]}-(\Theta/3)\,h_{ab}$
\ba\Theta \ &=& \ 3\, H + \CZ\, \dot T,\label{Theta}\\
 \sigma^a\,_b \ &=& \ \textrm{\bf{diag}}\,
[\,0,\,-2\Sigma,\,\Sigma,\,\Sigma],\label{sigma}\ea
where
\ba
\CZ \ &=& \
\frac{f+F+2\,f\,F\,T}{(1+f\,T)\,
(1+F\,T)},\label{CZ}\\
\Sigma \ &=& \ \frac{(F-f)\,\dot T}{(1+f\,T)\,
(1+F\,T)}.\label{Sigma}\ea
The metric (\ref{metric}) looks like a FLRW line element modified
by the terms containing $T,\,F$ and $f$. In fact, all $r$--dependent
variables derived above reduce to their FLRW forms:
$\rho=\sigma=0,\,\Theta/3= H$,\, if these ``pertubations'' vanish, {\it
i.e} if either $T=0$ or $f=f'=0$. This homogeneous subcase is a FLRW
spacetime whose source is the DE perfect fluid with matter--energy density and
pressure given by $\mu$ and $p$. In a sense, if $f\,T\ll 1 $ and $F\,T\ll 1$ the models
would correspond formally to specific exact perturbations of FLRW cosmologies.

\section{A time dependent $\Lambda$--field}

It is possible to interpret the homogeneous DE fluid as a time dependent ``cosmological
constant''. For this purpose we recall that (\ref{ltbmetric}) and (\ref{metric}) are
particular cases of the general spherically symmetric spacetime
\begin{equation}ds^2 = -A^2(t,r)\,c^2 dt^2
+B^2(t,r)\,dr^2+Y^2(t,r)\,d\,\Omega^2.\label{genss}\end{equation}
If the source of (\ref{genss}) is also
a perfect fluid like (\ref{TabPF}), we can define a ``mass function'' $M$
\begin{equation}M =
\frac{Y}{2}\,\left[1+\frac{\dot
Y^2}{A^2}-\frac{Y'{}^2}{B^2}\right],\label{Mdef}\end{equation}
satisfying
\bse\ba  2\,\dot M &=& -\kappa\,p\,Y^2\dot Y,\\
2\,M' &=& \kappa\,e\,Y^2\,Y'.\label{eqMr} \ea\ese
For spherical symmetry and in the appropriate limit, $M$ coincides with the ADM
and Hawking masses.  In particular, for the metric (\ref{ltbmetric}) in the form of
(\ref{metric}) and matter energy density (\ref{edesc}), $M$ can be found by integrating
(\ref{eqMr})
\begin{equation} 2\,M =
\kappa\int{(\rho
c^2+\mu)Y^2Y'\,dr}=2\,\M+\frac{\kappa\,\mu}{3}\,Y^3,\label{ADMmass}\end{equation}
with
\ba  2\,\M \ &\equiv& \ \kappa\,c^2\int{\rho Y^2Y'\,dr} \nonumber\\
&=&\frac{4}{9}\,R^3\,r^3 f\,\dot T\,[\,3\,H\,(1+f\,T)+f\,\dot
T\,],\label{cMdef}\ea
where we have assumed $M(t,0)=0$ and used $Y$ and $Y'$ given by (\ref{metric}), as well
as (\ref{eq_rho}) and (\ref{Fdef}) expressed as $F=2(r^{3/2}f)'/(3\sqrt{r})$. Equation
(\ref{Mdef}) yields then the Friedman--type evolution equation
\begin{equation} \dot Y^2 =
\frac{2\,M}{Y}=\frac{2\M}{Y}+\frac{\kappa\,\mu}{3}\,Y^2,\label{Freq1}\end{equation}
in which $\M$ and $\kappa\mu$ respectively play the roles of ``efective mass'' of the
inhomogeneous dust (DM) and a homogeneous time dependent $\Lambda$ term (DE). The mass
function $M$ can be related to a volume average of the total energy density by
\ba\kappa\,\la \rho c^2+\mu\ra &=& \frac{\kappa\int{(\rho
c^2+\mu)Y^2Y'\,dr}}{\int{Y^2Y'\,dr}}=\frac{6\,M}{Y^3}
\nonumber\\ &=& \frac{6\,\M}{Y^3}+\kappa\mu.\label{aver1}
\ea
Hence, we can identify the averaged DM and DE densities
\bse\label{aver2}\ba\kappa\,c^2\la \rho\ra &=& \frac{6\,\M}{Y^3} \ = \
\frac{4[\,3\,H\,(1+f\,T)+f\,\dot T\,]}{3\,(1+f\,T)^2}\,f\,\dot T,\nonumber\\\\
\kappa \la \mu\ra &=& \kappa\,\mu \ = \ 3\,H^2,\ea\ese
so that (\ref{Freq1}) can be related to (\ref{aver2})
and expressed as
\begin{equation} \frac{\dot Y^2}{Y^2} \ = \
\frac{\kappa}{3}\,\la\rho\ra\,c^2+\frac{\kappa}{3}\,\la\mu\ra,\label{Freq2}
\end{equation}
which looks like a FLRW Friedmann equation for a source made up with dust plus a time
dependent $\Lambda$ term, $\mu$, but both given in terms of their averaged densities
(\ref{aver2}). In this context, we can identify the relative velocity
$\dot Y/Y$ as a sort of averaged Hubble factor.

\section{Regularity conditions and asymptotics}

For a spherically symmetric spacetime (\ref{metric}) characterized by
(\ref{edesc})--(\ref{eq_eps}), the functions $R$,\,
$T$,\,$f$ and $F$ must comply with
\bse\label{regranges}\ba R \ > \ 0,\\ 1+f\,T \ >\ 0,\label{regY}\\
1+F\,T \ > \ 0.\label{regYr}\ea\ese
Hence, from (\ref{eq_rho}), the condition $\rho\geq 0$ becomes
\begin{equation} [\,4\,(\,3\,T\,H+\dot T)\,f\,F+6\,
H\,(f+F)]\,\dot T \ \geq \ 0\end{equation}
so that we can identify $t$ such that $R=0$ with a simultaneous big--bang
singularity associated with $\mu\to\infty$ (with $\rho$ finite) and two singularities
for which $\rho\to\infty$ (with $\mu$ finite)
\bse\ba 1+f\,T \ = \ 0,\label{singY0}\\
1+F\,T \ = \ 0,\label{singYr0}\ea\ese
which, in general, will be marked as  non--simultaneous surfaces in the $(t,r)$
coordinate plane. Notice that (\ref{singY0}) is a non--simultaneous big--bang,
or a central singularity ($\Rightarrow\,Y=0$), while (\ref{singYr0}) is a shell
crossing singularity ($\Rightarrow\,Y'=0$), both analogous to singularities in Lema\^\i
tre--Tolman--Bondi and Szekeres dust solutions~\cite{singSZ}. Hence, we will demand
that
\begin{equation} Y'>0 \,\, \textrm{holds for
all}\,\,(t,r)\,\,\textrm{complying with}\,\, Y>0,\label{noshellX}\end{equation}
or equivalently, that (\ref{singYr0}) does not occur in the coordinate range
that complies with (\ref{regY}). As shown in figure 1, condition (\ref{noshellX}) can be
be satisfied by suitable choices of free parameters.

\begin{figure}
\centering
\includegraphics[height=7cm]{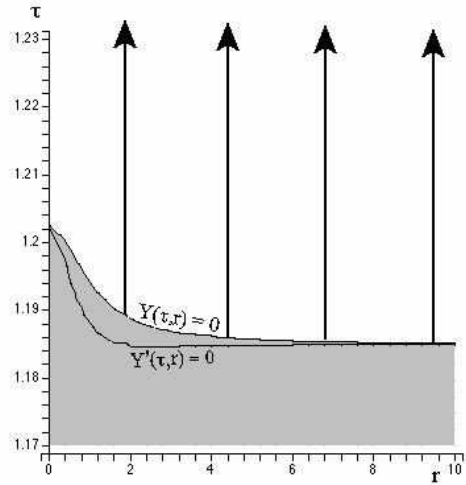}
%
%
\caption{ The plot depicts the coordinate representation of the non--simultaneous ``big
bang'' singularity $Y(\tau,r)=0$ as a ``surface'' in the $(\tau,r)$ plane, with the
dimensionless time $\tau$ defined by (55a) and other free parameters given by (54), (55)
and (60). The domain of regularity, $Y>0$, is shown with worldlines of fundamental
observers (arrows). Notice that the surface $Y'=0$ occurs in the region with negative $Y$
(gray shading).}
\label{fig1}       
\end{figure}

A usual regularity demand in spherically symmetric spacetimes is the existence of a
symmetry center, characterized by the regular vanishing of the surface area of the
orbits of SO(3). This condition defines the worldline of a privileged ``central''
observer for whom the universe appears isotropic. From
(\ref{metric}) and marking this worldline by
$r=0$, we must have for all $t$
\begin{equation} Y_{(c)}=\dot Y_{(c)}=0\end{equation}
where $Y_{(c)}=Y(t,0)$. We shall use henceforth the subindex ${}_{(c)}$ to denote
evaluation at $r=0$. Since spacial gradients of  $\rho,\,\Theta,\,\Sigma$ must vanish
at the center,  equations (\ref{eq_rho}) and (\ref{Fdef})--(\ref{Sigma}) imply that
$f_{(c)}$ must be bounded and $f'_{(c)} =0$, so that $F_{(c)}=f_{(c)}$. The
central values of these quantities are then given by
\bse\label{centralvals}\ba \kappa\,\rho_{(c)}\,c^2 &=& \frac{4}{3}\,f_{(c)}\,\frac{3\,
H\,[1+f_{(c)}\,T]+f_{(c)}\,\dot T}{[1+f_{(c)}\,T]^2}\,\dot
T,\\
\frac{\Theta_{(c)}}{3} &=& \left[\frac{\dot Y}{Y}\right]_{(c)}=H+\frac{2\,f_{(c)}\,\dot
T}{3\,[1+f_{(c)}\,T]},\\ \Sigma_{(c)} &=& 0\qquad \Rightarrow\quad
[\sigma^a\,_b]_{(c)}=0\ea\ese
where $f_{(c)}> 0$ and $f(r)$ for $r>0$ must be selected so that $\rho$ decreases
with increasing $r$.

In order to get an idea of the behavior of the models in the asymptotic limit
$t\to\infty$, we may assume as $t\to\infty$ an asymptotic power law scaling $R\to
t^k$ for $k>0$. From (\ref{eq_eps}), we have $\dot T\to 0$ in this limit, but $T$
tends to a finite asymptotic value ($T\to T^*$) only for $k>1/3$. From
(\ref{eq_mu}), (\ref{eq_eps}) we have
\begin{equation}\mu \ \to \ \frac{3\,k^2}{t^2},\qquad p \ \to \
\frac{k\,(2-3k)}{t^2}\end{equation}
while, assuming $f$ and $F$ everywhere finite, we have from (\ref{eq_mu})
and (\ref{eq_rho}) the ratio
\bse\ba\frac{\rho\,c^2}{\mu} &\to&
\ 0, \quad \textrm{for}\quad k>1/3,\label{k1}\\
\frac{\rho\,c^2}{\mu} &\to&
\ \textrm{const}, \quad \textrm{for}\quad 0<k<1/3\label{k2}.\ea\ese
Therefore, the  homogeneous DE fluid dominates asymptotically over the DM component 
even if $p>0$ (or equivalently $k<2/3$), while values of $k$ associated with
negative $p$ always yield the asymptotic ratio (\ref{k1}). It is not difficult to
verify that the same asymptotic behavior occurs if $R$ follows exponential or
logarithmic asymptotic scalings, $R \to\exp(k\,t)$ or $R\to\ln\,t$. Thus, it is
generically possible to have an asymptotically  DE dominated asymptotic scenario for all
fundamental observers, which is an important property of the models under
consideration. As long as $R$ scales asymptotically faster than $R\to t^{2/3} $ and
\,$f,\,F$, are finite everywhere, we have $p<0$ together with
\begin{equation}\frac{\Theta}{3}\to H,\qquad  \Sigma\to
0,\qquad \textrm{as}\quad t\to\infty,\end{equation}
so that, regardless of the choice of $f$ (spacial dependence), the models homogenize and
isotropize for all fundamental observers ($R$ constant) as $t\to\infty$, implying that
all fundamental observers detect in this asymptotic limit local conditions that are very
close to those of a FLRW cosmology characterized by a DE source with $\mu$ and $p$.

It is also interesting to examine the behavior of the models as $r\to\infty$ along
hypersurfaces of constant $t$, the rest frames of fundamental comoving observers. It is
important to notice that the radial coordinate, $r$, merely labels the fundamental
comoving observers, so it has no invariant meaning and can be though of as a dimensionless
ratio, such as $r=\ell/\ell_0$, where $\ell_0$ is the present value of an arbitrary length
scale (for example, the scale of homogeneity $\ell_0\sim 100-300$ Mpc). Thus, we can
consider ``local'' scales ($\ell \ll \ell_0$) as marked by
$r\ll 1$, so that asymptotic ``cosmic'' scales correspond to $r\gg 1$, while the
``transition'' from local to cosmic scales is roughly given by $r\sim 1$. 

Since $f$ is the free function that governs the dependence on $r$, the variation of all
quantities in the radial direction is tied to specific choices of this free function. In
particular
$f$ should be selected so that $\rho$ decays as $r$ increases. Also, the spacial
dependence of $\M$, the ``effective mass'' of DM defined by (\ref{cMdef}), is governed by
the term
$r^3f$, thus it is also convenient to demand that
$f\geq 0$ and that $r^3f$ must be a monotonously increasing function. Still, considering
all these  restrictions, it is possible to select $f$ so that
\begin{equation} f \to 0,\quad F\to 0,\qquad \textrm{as}\quad
r\to\infty,\label{infvals01}\end{equation}
which is a sufficient condition for
\begin{equation}\frac{1}{3}\,\Theta|{}_{r=\infty}\to H,\qquad \rho|{}_{r=\infty}\to
0,\qquad \Sigma|{}_{r=\infty}\to 0.\label{infvals02}\end{equation}
Thus, since $\mu$ is independent of $r$, these limits imply that local conditions
for fundamental observers with large $r$ are similar those of fundamental observers
of a FLRW cosmology whose source is a fluid with $\mu$ and $p$ (that is, a ``pure'' DE
fluid). Another possibility is furnished by the choice:
\begin{equation} f \to f^*,\quad F\to f^*,\qquad \textrm{as}\quad
r\to\infty,\label{infvals11}\end{equation}
where $f^*<f_{(c)}$ is a positive constant. This yields
\bse\label{infvals12}\ba \kappa\,\rho|{}_{r=\infty}\,c^2 &=&
\frac{4}{3}\,f^*\,\frac{3\, H\,[1+f^*\,T]+f^*\,\dot
T}{[1+f^*\,T]^2}\,\dot T,\\
\frac{1}{3}\,\Theta|{}_{r=\infty} &=& \left[\frac{\dot
Y}{Y}\right]_{r=\infty}=H+\frac{2\,f^*\,\dot T}{3\,[1+f^*\,T]},\\
\Sigma|{}_{r=\infty} &=& 0\qquad \Rightarrow\quad [\sigma^a\,_b]_{r=\infty}=0.\ea\ese
The preference of one of these choices of asymptotic behavior along the rest
frames depends on the problem one is interested to study: if we want to
examine a large scale (supercluster scale or larger) spherical inhomogeneity whose
evolution requires that we somehow ``plug in'' the effects of a cosmological background,
then the choice (\ref{infvals11}) may be preferable, while the choice
(\ref{infvals01}) may be preferable for a relatively small scale and/or
large density contrast description of an homogeneity (cluster of galaxies) that
ignores cosmic effects (see figure 3).

It is also possible to perform a smooth matching, for a comoving
radius $r_b$, of a region $0\leq r\leq r_b$ of the DM and DE
mixture to a spatially flat FLRW spacetime characterized by $\mu$
and $p$, occupying $r>r_b$. Necessary and sufficient conditions
follow if $f_{(b)}=F_{(b)}=0$, where ${}_{(b)}$ denotes evaluation
at $r=r_b$, so that $\rho_{(b)}=\Sigma_{(b)}=0$ and
$\Theta_{(b)}/3=H$ hold for the FLRW region. However, such a
matching also requires the mass function $M$ in (\ref{Mdef}) to be
continuous (at least $C^0$) at $r=r_b$, which from (\ref{ADMmass})
implies $\M_{(b)}=0$, but $\M$ is a volume integral of $\rho$.
Therefore, if this integral must vanish at $r=r_b$ for all $t$,
the dust density $\rho$ in the integrand must necessarily be
negative in a finite domain of the mixture region $0\leq r\leq
r_b$. Hence, we will not consider this type of matching any further.

\section{Observational parameters}

The quantity $H$ in (\ref{Hhat}) is the Hubble expansion factor
associated with a FLRW geometry, for the inhomogeneous metric (\ref{metric}) the
proper generalization of this parameter is given by~\cite{ellis,HMM}
\begin{equation}\CH \ = \ \frac{\Theta}{3}+\sigma_{ab}\,n^a
n^b,\label{H1}\end{equation}
where the vector $n^a$ complies with $n_an^a=1,\,u_an^a=0$. For
a spherically symmetric spacetime, it is necessary to evaluate $n^a$ for
general comoving observers located in an ``off--center'' position in the
spherical coordinates $(r,\theta,\phi)$ centered at $r=0$. For the metric (\ref{metric})
equation (\ref{H1}) becomes in general
\ba \CH \ &=& \
 H + \F\,\dot T\nonumber
\\ \F \ &=& \
\frac{2\,[f\,(1+F\,T)+\frac{3}{2}\,(F-f)\,\cos^2\psi]}
{3\,(1+f\,T)\,(1+F\,T)},\label{Hdef}
\ea
where $\psi$ is ``observation angle''
between the direction of a light ray and the ``radial'' direction for a fundamental
observer located in $(r,\theta,\varphi)$~\cite{HMM}.  Therefore, the exact local
values of the observational parameters $\Omega$ for DE and DM are
\ba \Omega_{\textrm{\tiny{DE}}} \ &=& \ \frac{\kappa\,\mu}{3\,\CH^2} =
\frac{H^2}{\left[H+\F\,\dot T\right]^2},\label{OMDE}\\
\Omega_{\textrm{\tiny{DM}}} \ &=& \ \frac{\kappa\,\rho\,c^2}{3\,\CH^2}  =
\frac{\kappa\,\rho\,c^2}{ 3\,\left[H+\F\,\dot T\right]^2}  =
\frac{\rho\,c^2}{\mu}\,\Omega_{\textrm{\tiny{DE}}},\label{OMDM}
\ea
while the acceleration parameter is~\cite{ellis}
\begin{equation}q \ = \
\frac{6\Sigma^2}{\CH^2}+\frac{\Omega_{\textrm{\tiny{DE}}}
+\Omega_{\textrm{\tiny{DM}}}}{2}\,\left[1+\frac{3\,p/\mu}{1+\rho
c^2/\mu}\right],\label{qdef}\end{equation}
where $\Sigma$ is given by (\ref{Sigma}).

If we consider the flow of cosmic DM with density $\rho$ at the
length scale of the observable universe ($\sim 3000\,h$ Mpc), then
present day values of shear and DM density gradients in comparable scales are severely
restricted by the near isotropy of the CMB~\cite{CMBinh}
\ba\left[\frac{|\sigma_{ab}\,\sigma^{ab}|^{1/2}}{\Theta}\right]_0 \ = \
\left[\frac{\sqrt{6}\,|\Sigma|}{\Theta}\right]_0
\ \alt \ 10^{-5},\label{bound1}\\
\left[\frac{h_a^b\,\rho_{,b}}{\rho}\right]_0 \ = \
\left[\frac{\rho\,'}{\rho}\right]_0 \ \alt
\ 10^{-5}.\label{bound2}\ea
Hence the large scale spacial dependence of the observational parameters
(\ref{H1}), (\ref{OMDE}), (\ref{OMDM}) and (\ref{qdef}) must also
be restricted by these bounds. However, these restrictions can be
strongly relaxed, at a local level, if we examine the spacial variation of local values of
DM density and observational parameters in scales smaller than the homogeneity scale
$\sim 100-300$ Mpc. As mentioned in the previous section, scale considerations influence
the choice of boundary conditions (\ref{infvals01}) or (\ref{infvals11}). 

\begin{figure}
\centering
\includegraphics[height=12cm]{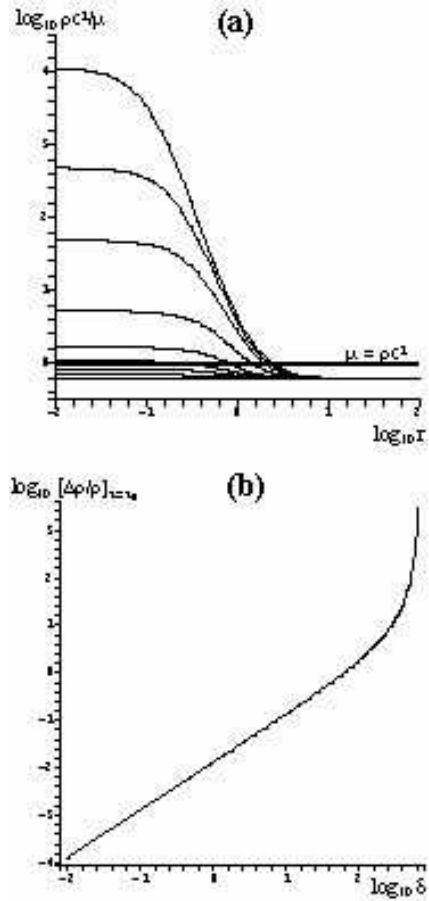}
%
%
\caption{ Figure (a) illustrates how the present DM density profile,
$\rho/\mu$, exhibits an increasingly large
inhomogeneous contrast as $\delta$ ranges from $\delta=670$ (maximal contrast) to
$\delta=0.1$ (minimal contrast). We have used (54) and (55) and the free parameters
$f,\,\gamma,\,f^*,\,T^*$ are those given by (60) and (64). Notice how for all $\delta$,
we have $\mu>\rho c^2$ in the cosmic background region, with all curves tending as
$r\to\infty$ to the cosmic ratio
$\Omega_{\textrm{\tiny{DM}}}/\Omega_{\textrm{\tiny{DE}}} \approx 0.56$. Figure (b)
displays the plot of a measure of density contrast between the center and
$r=\infty$, given by $\Delta\rho/\rho=\rho(\tau_0,0)/\rho(\tau_0,\infty)-1$, as a
function of $\log_{10} \delta$. Notice how the contrast grows proportionally to
$\delta$, well beyond the linear perturbation regime.}
\label{fig2}       
\end{figure}  

\begin{figure}
\centering
\includegraphics[height=12cm]{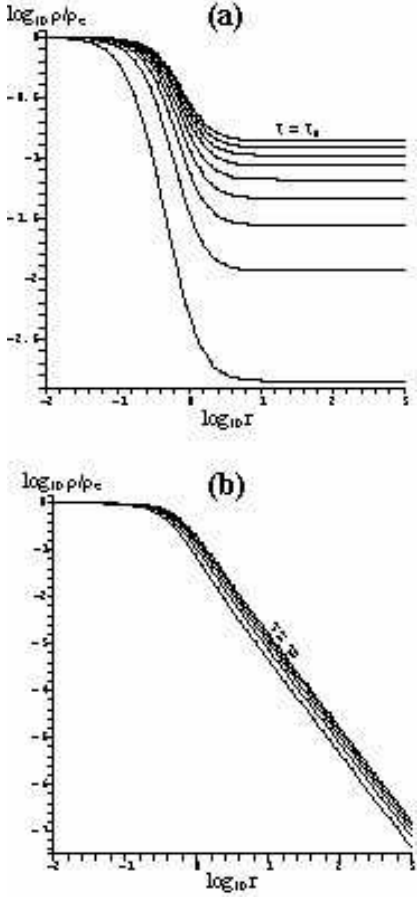}
%
%
\caption{ The panels depict the logarithmic profile of $\rho/\rho_{(c)}$, with
$\rho_{(c)}=\rho(\tau,0)$ for a sequence of values of constant $\tau$, up to the
present $\tau=\tau_0$. The free parameters $\gamma,\,\delta,\,T^*$ are those given by
(64). In panel (a) we selected $f^*=100$, hence $\rho/\rho_c$ tends to the
nonzero asymptotic cosmic DM density given by (46a). We see a power law decay of DM
density only in the transition ($r\sim 1$) from the central overdensity region to the
cosmic background. In panel (b) we selected $f^*=0$, thus the curves closely resemble
isothermal profiles, in which $\rho$ decays to zero approximately as $r^{-2}$. This is
not surprising since $\M\to r$ asymtotically with the choice (60).}
\label{fig3}       
\end{figure}

\section{A simple example: the ``gamma law''.}

In order to illustrate how to work out the expressions we have derived and how to
calculate relevant quantities of the models, we consider now the simple case of a
homogeneous DE fluid satisfying a simple equation of state known as the ``gamma--law''
\begin{equation} p \ = \ (\gamma - 1)\,\mu,\label{glaw}
\end{equation}
where $\gamma$ is a constant. The dust plus homogeneous fluid
mixtures that we are studying were examined
previously~\cite{old2F1}, assuming (among other choices) this
equation of state, but placing especial emphasis in a dust and
radiation ($\gamma=4/3$) mixture. Since our emphasis is now on
modelling DE sources, we will assume $0 < \gamma <2/3$,
so that $-1<p/\mu<-1/3$. In this case we have from (\ref{eq_p}),
(\ref{eq_eps}), (\ref{eq_mu}), (\ref{Hhat}) and (\ref{Hdot})
\bse\label{glawpars}\ba a &\equiv& \frac{R}{R_0} =
\tau^{2/3\gamma},\qquad \tau \ = \
\frac{3}{2}\,\gamma\,h\, H_0\, t,\\
\kappa\,\mu  &=&  3\,H^2 \ = \ \frac{3\,(hH_0)^2}{\tau^2},\label{eq_mu1}\\
T &=& T^* - \frac{\gamma_1}{\tau^{1/\gamma_1}},\qquad
\gamma_1 \ = \ \frac{\gamma}{2-\gamma}\ea\ese
where $ H_0 =100\,\textrm{km}/(\textrm{sec\,Mpc})$,\,$h=0.7$ and
$T^*$ is a dimensionless constant denoting the asymptotic
value of $T$ (we have then in (\ref{eq_eps}) the choice
$c_0=\frac{3}{2}\,\gamma\,h$).

Notice that the assumptions (\ref{glawpars}) have been obtained
from the FLRW equations (\ref{eq_p}) and (\ref{eq_mu}) and yield a
power law form for the function $a$ (equivalent to the FLRW scale
factor). Therefore, following~\cite{CJSF}, this form of the
homogeneous DE fluid is equivalent to a scalar field with an
exponential potential following the so--called ``scaling law''.

The density of the dust component and the generalized Hubble
factor are found by inserting (\ref{glawpars}) into (\ref{eq_rho})
and (\ref{Hdef})
\begin{widetext}
\ba\kappa\rho c^2 \ &=& \ \frac{3\,\gamma^2\,
H_0^2}{\tau^2}\,\frac{\left[2\,T^*\,f\,F+(1/\gamma)\,(f+F)\right]\,
\tau^{1/\gamma_1}+\gamma_2\,f\,F}{\left[\,(1+f\,T^*)\,
\tau^{1/\gamma_1}-\gamma_1\,f\,\right]\,\left[\,(1+F\,T^*)\,
\tau^{1/\gamma_1}-\gamma_1\,F\,\right]},\qquad \gamma_2 \
= \ \frac{2-3\,\gamma}{2-\gamma},\label{eq_rho1}\\
\CH \ &=& \ \frac{h\,H_0}{\tau} \, \left\{1+ \gamma\,\frac{\left[\,(1+F\,T^*)\,
\tau^{1/\gamma_1}-\gamma_1\,F\,\right]
f+\frac{3}{2}\,(F-f)\,\tau^{1/\gamma_1}\,\cos^2\psi}{\left[\,(1+f\,T^*)\,
\tau^{1/\gamma_1}-\gamma_1\,f\,\right]\,\left[\,(1+F\,T^*)\,
\tau^{1/\gamma_1}-\gamma_1\,F\,\right]}\right\}.\label{eq_HH1}\ea
\end{widetext}
\begin{figure}
\centering
\includegraphics[height=6cm]{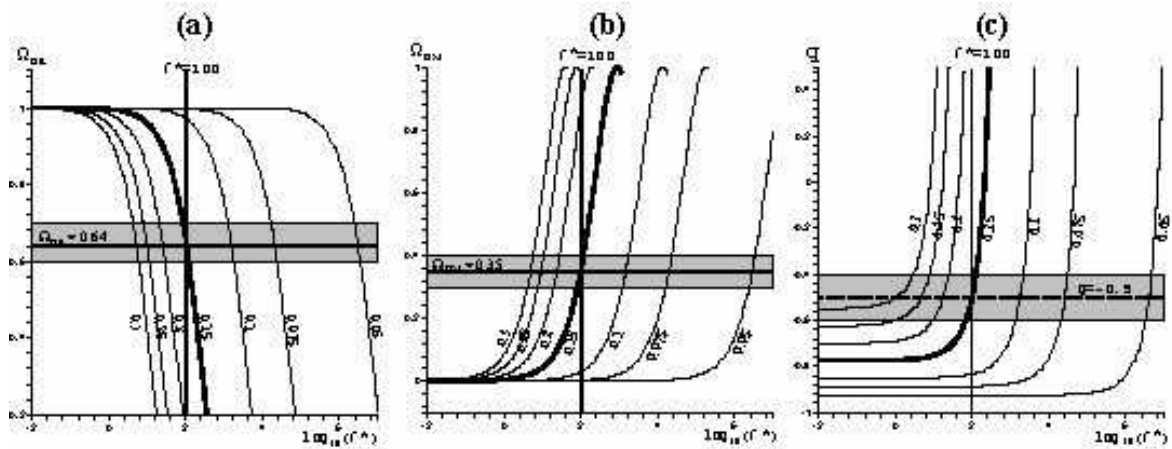}
%
%
\caption{ The asymptotic value of $f$ in (60) is given by $f^*$. Assuming $T^*=0$ and an
arbitrary $\delta$, we can find a suitable value for $f^*$, for any given 
$\gamma$ in (54), by demanding that the observational parameters
$\Omega_{\textrm{\tiny{DE}}},\,\Omega_{\textrm{\tiny{DM}}},\,q$ (panels (a),(b) and (c))
have appropriate ``present'' cosmological values in the ranges (63) as $r\to\infty$ for
$\tau=\tau_0$ (gray stripes). Each curve is marked by a given value of $\gamma$. Notice
that for $\gamma=0.15$, the choice $f^*\approx100$ yields
$\Omega_{\textrm{\tiny{DE}}}=0.64,\,\Omega_{\textrm{\tiny{DM}}}=0.35,\,q=-1/2$, which are
reasonably close to currently accepted observational data. For $\gamma$ closer to $0$
(cosmological constant), we would have to select larger values for $f^*$, while larger 
$\gamma$ close to $0.3$ correspond to $f^*\sim 1$.}
\label{fig4}       
\end{figure}
From (\ref{glawpars}), we see that $R$ scales as $t^k$, so that $k>1/3$ for
$\gamma<2$, hence for the $\gamma$ values that we are interested we should obtain the
ratio $\rho c^2/\mu$ given by (\ref{k1}). Using (\ref{eq_mu1}) and (\ref{eq_rho1}) and
assuming an arbitrary but finite $f$ and $F$, we obtain in the limit $\tau \gg 1$
\ba \frac{\rho\,c^2}{\mu}  &\to&
\frac{2\,T^*\,f\,F+(1/\gamma)\,(f+F)}{(1+f\,T^*)\,(1+F\,T^*)} \
\frac{3\,\gamma^2}{\tau^{1/\gamma_1}}
\ \to \ 0,\\
\CH &\to& \frac{h\,H_0}{\tau} = H\ea
indicating that for all cosmic observers the mixture homogenizes and isotropizes as the
homogeneous DE fluid dominates asymptotically over the cold DM component. 

\section{Numerical exploration.}

   Having found $T,\,H$ and $a=R/R_0$ for the particular case of a gamma law (\ref{glaw}),
we only need to select the function $f=f(r)$ in order to render the models fully
determinate.  A convenient form for $f$ is
\begin{equation} f \ = \ f^* + \frac{\delta}{1+r^2},\label{fform}\end{equation}
so that both type of asymptotic boundary conditions, (\ref{infvals01})--(\ref{infvals02})
or (\ref{infvals11})--(\ref{infvals12}), can be accomodated by selecting $f^*$ to be zero
or nonzero. Notice also that asymptotically as $r\to\infty$, we have:\, $\M\to r$ if
$f^*=0$ and $\M\to r^3$ if $f^*>0$.

Considering (\ref{regranges}), the regular evolution range for the models is the coordinate
range $(\tau,r)$ where (\ref{noshellX}) holds with $a,\,H,\,T$ given by (\ref{glaw}),
(\ref{glawpars}) and
$f,\,F$ given by (\ref{Fdef}) and (\ref{fform}). Thus, assuming as the initial
``big--bang'' the coordinate surface (\ref{singY0}) marking $Y=0$, we can take the big
bang time as the value $\tau=\tau_{\textrm{bb}}$ at this surface corresponding to
$r\to\infty$ (see figure 1), that is:
\begin{equation}\tau_{\textrm{bb}}^{1/\gamma_1}=\frac{\gamma_1\,(f^*+\delta)}{1+(f^*+\delta)\,T^*}\end{equation}
Thus, considering the ``age of the universe'' roughly as $\Delta t_0\approx 14$ Gys and
$h\approx 0.7$, the present cosmic era corresponds to
\begin{equation}\tau_0= \tau_{\textrm{bb}}+\frac{3}{2}\gamma\,h\,H_0\,\Delta
t_0\approx \tau_{\textrm{bb}}+3.17\,\gamma\end{equation}
As shown in figure 1, the big bang surface (\ref{singY0}) is not simultaneous, thus for
any hypersurface $\tau=$ constant, the regions near the center at $r=0$ will be
``younger'' than those asymptotically far at large values of $r$. 

The parameter $\delta=f_{(c)}-f^* >0$ in (\ref{fform}) provides the a measure of
inhomogeneity contrast, or ``spacial'' variation of all quantities along the rest frames
($t=$ constant) between the symmetry center $r=0$ and $r\to\infty$. Thus,  a
sufficiently large/small value of $\delta$ makes the values at $r=0$ and $r\to\infty$
sufficiently close/far to each other, thus indicating small/large ``contrast'' or degree
of inhomogeneity. We test the effect of $\delta$ along the present day hypersurface
$\tau=\tau_0$ by plotting in figure 2a the spacial profiles of $\rho c^2/\mu$ for a
sequence of $\delta$ values, while figure 2b depicts the density contrast between the
overdensity at center and the cosmological background, as a function of
$\log_{10}\delta$.

The effect of selecting $f^*$ zero or nonzero is shown in figures 3a and 3b, by
means of a comparison between the resulting DM density profiles that follow by evaluating
(\ref{eq_rho1}) and (\ref{fform}) along a sequence of hypersurfaces of  constant $\tau$
(up to $\tau_0$). The asymptotic condition (\ref{infvals01}) ($f^*=0$) yields an
approximately $r^{-2}$ power law decay of $\rho$ that
resembles a standard isothermal profile, all of which is consistent with
(\ref{infvals02}) and with the fact that the form of $f$ in (\ref{fform}), with
$f^*=0$, leads to $\M\to r$, the same asymptotic behavior of the standard ``isothermal
sphere''. On the other hand, the choice (\ref{infvals11}) ($f^*> 0$) leads also to a
power law decay of $\rho$, but towards a cosmological background with asymptotic density
given by (\ref{infvals12}).

The appropriate numerical value for the asymptotic constant, $f^*$, can be found by
demanding that the cosmological observational parameters
$\Omega_{\textrm{\tiny{DE}}},\,\Omega_{\textrm{\tiny{DM}}},\,q$, evaluated in the
cosmic background ($r\to\infty$) at the present era ($\tau=\tau_0$),  take (for a given
$\gamma$) reasonably close values to those currently accepted from observational data.
Since (\ref{eq_HH1}) with $T^*=0$ and $f$ given by (\ref{fform}), evaluated at
$\tau=\tau_0$ and $r\to\infty$, is independent of $\delta$ and $\psi$, we can plot
$\Omega_{\textrm{\tiny{DE}}},\,\Omega_{\textrm{\tiny{DM}}},\,q$ as functions of $f^*$ and
$\gamma$. As figures 4a, 4b and 4c illustrate, the desired value of $f^*$ for any given
$\gamma$ can be selected so that the forms (\ref{OMDE}), (\ref{OMDM}) and (\ref{qdef}) at
$\tau=\tau_0$ and $r\to\infty$ yield:
\bse\label{par_vals}\ba
0.6&<&\Omega_{\textrm{\tiny{DE}}}<0.7,\\
0.3&<&\Omega_{\textrm{\tiny{DM}}}<0.4,\\
-0.5 &<& q < -0.4\ea\ese
In particular, if we select $\gamma=0.15$, an appropriate value is $f^*\approx 100$,
leading to $\Omega_{\textrm{\tiny{DE}}}\approx 0.64$,\, $\Omega_{\textrm{\tiny{DM}}}
\approx 0.35$ and $q\approx -0.5$. Notice
that $\Omega_{\textrm{\tiny{DE}}}+\Omega_{\textrm{\tiny{DM}}}\approx 1$, 
but the present Omega for DM would be slightly higher than the currently accepted value
$\Omega_{\textrm{\tiny{DM}}}\approx 0.3$. Thus, we could argue that these parameter
values would become a very accurate approximation to actually inferred cosmological
parameters if we would consider $\rho$ as the compound density of DM and baryonic matter.
\begin{figure}
\centering
\includegraphics[height=6cm]{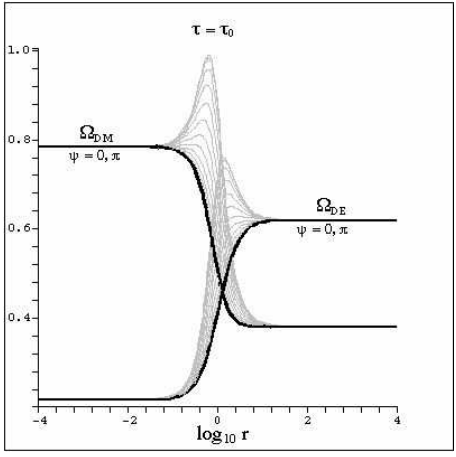}
%
%
\caption{ The figure illustrates the effects of anisotropy through the dependence of
the profiles of $\Omega_{\textrm{\tiny{DE}}}$ and $\Omega_{\textrm{\tiny{DM}}}$, along
$\tau=\tau_0$, on the off--center ``observation'' angle $\psi$. Free parameters are given
by (64).  The thick black curves correspond to $\psi=0,\pi$, while the gray curves provide
the profiles of $\Omega_{\textrm{\tiny{DE}}}$ and $\Omega_{\textrm{\tiny{DM}}}$ for
various other values of $\psi$. In particular, the larger ``peaks'' of these parameters
occur for $\psi=\pi/2,3\pi/2$ and around $r\sim 1$. Notice how these effects of anisotropy
are negligible in the overdensity region around $r=0$ and in the cosmic background for
large $r$. For the deceleration parameter $q$ these effects are negligible for all $r$.}
\label{fig5}       
\end{figure}

\begin{figure}
\centering
\includegraphics[height=6cm]{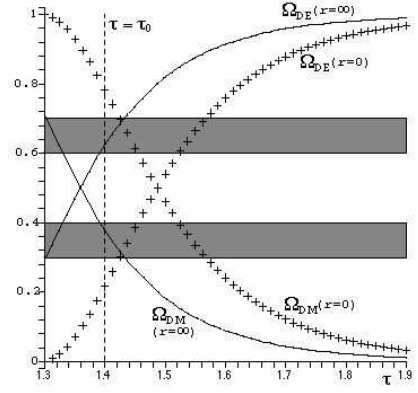}
%
%
\caption{ The figure illustrates the effects of inhomogeneity by showing how the
time evolution of $\Omega_{\textrm{\tiny{DE}}}$ and $\Omega_{\textrm{\tiny{DM}}}$ (for
$\psi=0$) is very different along the center of the overdensity at $r=0$ (crosses) and in
the cosmic background ($r\to \infty$, solid curves). The free parameters are given by
(64). Notice how in the present cosmic era ($\tau=\tau_0$) DE dominates over DM in the
cosmic background, but DM dominates over DE in the center.}
\label{fig6}       
\end{figure}

\begin{figure}
\centering
\includegraphics[height=14.5cm]{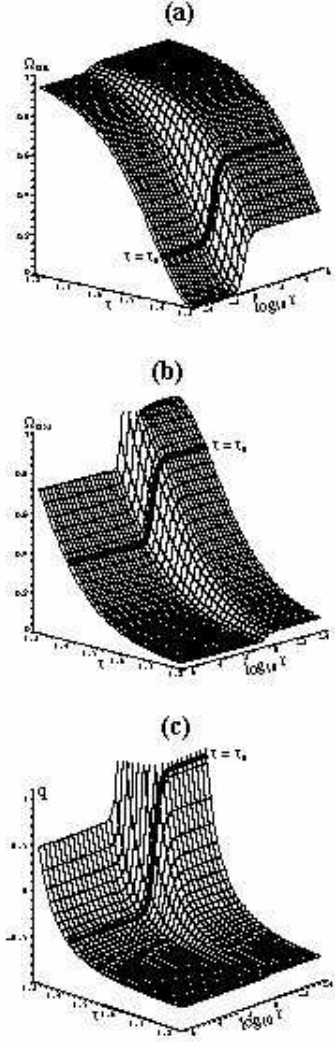}
%
%
\caption{ The figures illustrate the full dependence of
$\Omega_{\textrm{\tiny{DE}}},\,\Omega_{\textrm{\tiny{DM}}}$ and
$q$ on $\tau$ and $r$ (panels (a), (b) and (c)) for $\psi=0$ and with free parameters
given by (64). The thick black curves denote the hypersurface of present cosmic time
$\tau=\tau_0$. The figures clearly show the differences among the overdense region around
$r=0$, the cosmic background for large $r$ and the transition zone between them around
$r\sim 1$. As in the previous figure, it is evident that DE dominates over DM in the
cosmic background, but DM dominates over DE in the center, with both converging as
$\tau\to\infty$ to asymptotically homogeneous states with values
$\Omega_{\textrm{\tiny{DE}}}\to 1,\,\Omega_{\textrm{\tiny{DM}}}\to 0$. Notice in (c) how
the deceleration parameter, $q$, is negative in the cosmic background (accelerating
universe), but is positive in the overdense region where the dynamics of local gravity
should not be repulsive.}
\label{fig7}       
\end{figure}

\begin{figure}
\centering
\includegraphics[height=6cm]{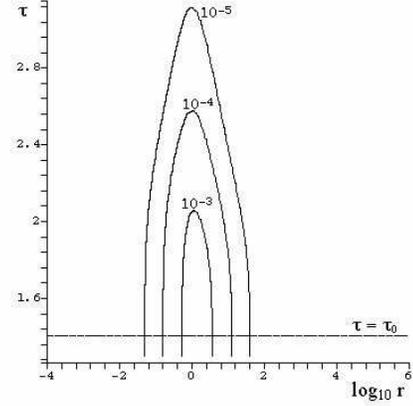}
%
%
\caption{ Level curves of the ratio $\sqrt{6}\Sigma/\Theta$. Free parameters are
those given by (64). Notice that condition (52) is satisfied for most of the range of $r$
at $\tau=\tau_0$, failing to hold only in the transition ($r\sim 1)$ between the
overdensity and the cosmic background.}
\label{fig8}       
\end{figure}

\begin{figure}
\centering
\includegraphics[height=6cm]{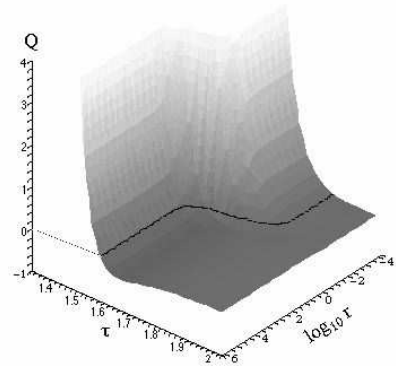}
%
%
\caption{ The interaction function $Q(\tau,r)$ given by (66). Free parameters are given by
(64) and the level curve $Q=0$ is the thick black curve. Notice how $Q>0$ in the present
cosmic era (DE transfers energy to DM), then changes sign (DM transfers energy to DE),
with $Q\to 0$ for large $\tau$, thus indicating an asymptotic state in which DE and DM
only interact gravitationally. }
\label{fig9}       
\end{figure}
 
\begin{figure}
\centering
\includegraphics[height=6cm]{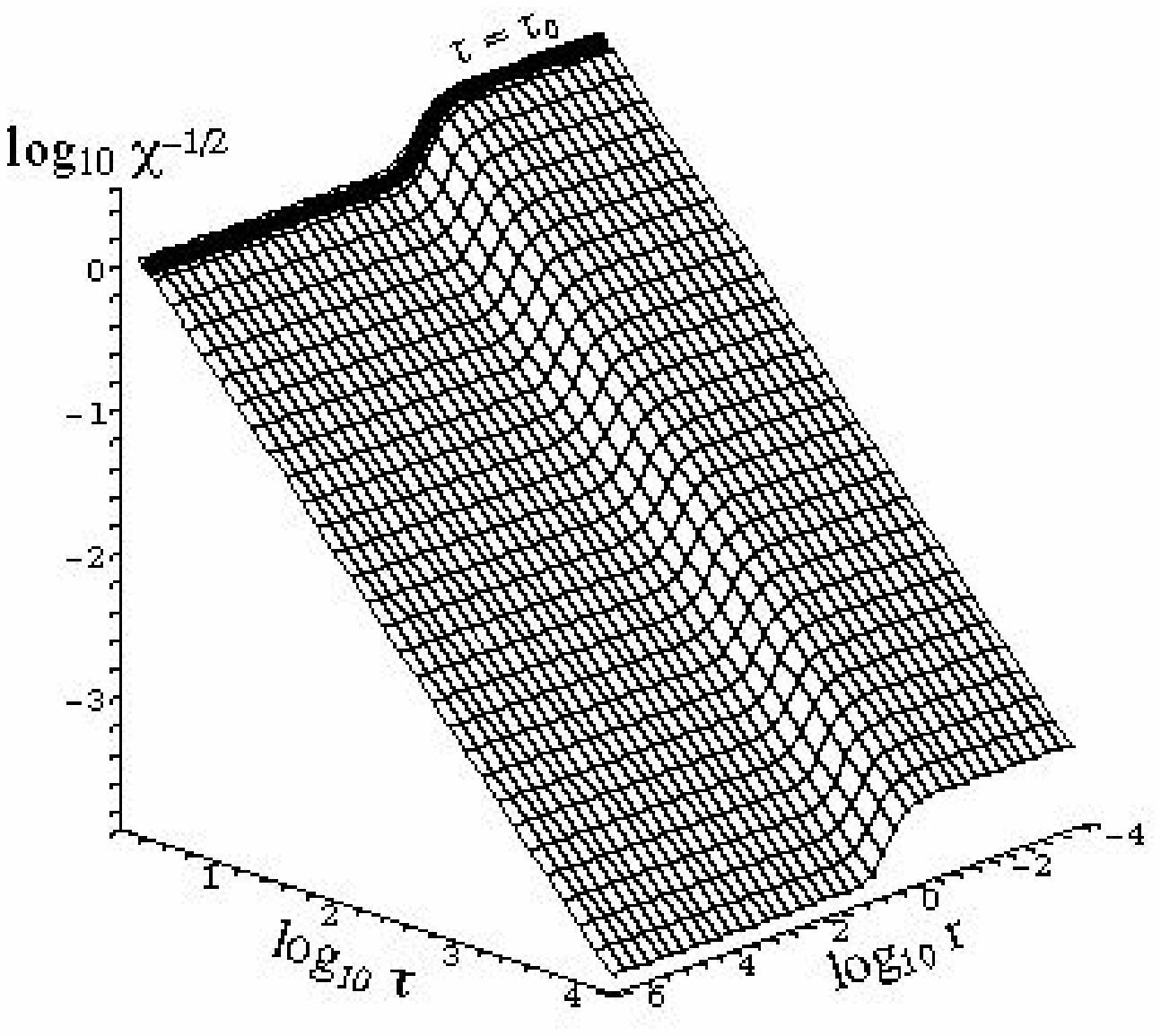}
%
%
\caption{Qualitative logarithmic plot of the coupling function $\chi^{-1/2}$ defined by
(71), with $\xi\propto 1/r^2$ and free parameters given by (64). The hypersurface
$\tau=\tau_0$ is displayed as a thick black curve. Notice the clear power law time
dependence of $\chi^{-1/2}$, just slightly modulated by the spacial dependence. This
indicates that $\chi=\chi(\phi)$ holds approximately for a scalar field associated with
(54) and (67). }
\label{fig10}       
\end{figure}

The efects of anisotropy emerge in the dependence of $\CH$, as given by (\ref{eq_HH1}), on
the off--center  ``observation angle'' $\psi$. This implies dependence on $\psi$ for
$\Omega_{\textrm{\tiny{DE}}},\,\Omega_{\textrm{\tiny{DM}}},\,q$. Considering the free
parameter values
\begin{equation}T^*=0,\quad f^*=100,\quad \delta=200,\quad
\gamma=0.15,\label{freepars}\end{equation}
complying with the cosmic background ranges (\ref{par_vals}), figure 5 displays
$\Omega_{\textrm{\tiny{DE}}}$ and $\Omega_{\textrm{\tiny{DM}}}$, evaluated at
$\tau=\tau_0$, as a function of $\log_{10} r$ for assorted fixed values of $\psi$. The
same profiles of $\Omega_{\textrm{\tiny{DE}}}$ and
$\Omega_{\textrm{\tiny{DM}}}$ occur for $\psi=0$ and $\psi=\pi$ (thick black lines) and,
in general for any two values of $\psi$ that differ by a phase of $\pi$, with the
highest ``peaks'' corresponding to gray curves with $\psi=\pi/2$ and $\psi=(3/2)\pi$. This
singles out two ``preferential'' distinctive directions: one along the axis
$\psi=0,\,\pi$ and the other along $\psi=\pi/2,\,(3/2)\pi$. This is a clear
representation of a quadrupole pattern, as expected for a geodesic but shearing
4-velocity~\cite{ellis,HMM}. Also, as revealed by figure 5, the curves for the various
$\psi$ differ from each other only in the transition region near
$r\sim 1$, thus the effects of this quadrupole anisotropy are negligible near the center
of the local overdensity and in the cosmic background asymptotic region.  

The effects of inhomogeneity are illustrated by figure 6, for the
same parameters (\ref{freepars}) but keeeping $\psi=0$ fixed, by plotting
$\Omega_{\textrm{\tiny{DE}}}$ and $\Omega_{\textrm{\tiny{DM}}}$ as a function of $\tau$
for $r=0$ (crosses) and $r=\infty$ (solid lines). Notice how for the present cosmic era,
$\tau=\tau_0$,  we have $\Omega_{\textrm{\tiny{DE}}}$ dominating at the cosmic background
region ($r\to\infty$), but $\Omega_{\textrm{\tiny{DM}}}$ dominates in the overdensity
region ($r=0$). This effect of inhomogeneity can be further appretiated in figures
7a, 7b and 7c, displaying $\Omega_{\textrm{\tiny{DE}}}$,\,
$\Omega_{\textrm{\tiny{DM}}}$ and $q$ as functions of $\tau$ and $\log_{10} r$ for
$\psi=0,\,\pi$ and using the free parameters (\ref{freepars}). It is particularly
interesting to remark how as $\tau$ grows we have: \, $\Omega_{\textrm{\tiny{DE}}}\to
1$,\,$\Omega_{\textrm{\tiny{DM}}}\to 0$ and $q\to -1/2$ for all $r$, as expected for a  DE
dominated asymptotically future scenario and associated with an ever accelerating universe
that follows a ``repulsive'' dynamics. However,  in the present cosmic era (thick black
curve) this repulsive accelerated dynamics on which $\Omega_{\textrm{\tiny{DE}}}$
dominates and $q<0$ only happens in the cosmic background region, with
$\Omega_{\textrm{\tiny{DM}}}>\Omega_{\textrm{\tiny{DE}}}$ and $q>0$ (\textit{i.e.}
``attractive'' dynamics) in the local overdensity region with a relatively
large DM density contrast $\delta=200$. 

Conditions (\ref{bound1}) and (\ref{bound2}) place stringent limits on large
scale deviations from homogeneity and anisotropy, but these bounds do not apply to local
values of these quantities. It is interesting to plot these quantities for the cases
depicted in the previous figures, all characterized by (\ref{glaw}), (\ref{glawpars}),
(\ref{fform}) and (\ref{freepars}). Figure 8 depicts level curves of the logarithm of the
shear to expansion ratio of (\ref{bound1}), with $\Theta$ and $\Sigma$ given by
(\ref{Theta})--(\ref{Sigma}). Even considering the relatively large value $\delta=200$,
condition (\ref{bound1}) holds througout most of the coordinate range $(\tau,r)$
including the far range ``cosmological background'' region of large $r$ and the local
overdensity region near the symmetry center $r=0$, so that for the present cosmic time
$\tau=\tau_0$ it only excludes the relatively small scale local region around $r\sim 1$
that marks the ``transition'' from the local overdensity to the cosmic background.
However, a choice  like $\delta=0.01$ would yield similar level curves, but with values
three orders of magnitude smaller, thus denoting a state of  almost global homogeneity,
since (\ref{bound1}) would hold in almost all local scales in
$\tau=\tau_0$. A graph that is qualitatively very similar to that of figure 8 emerges for
condition (\ref{bound2}).

This difference between
the dynamics of local inhomogeneities and that of the cosmic background cannot
be appretiated in such a striking and spectacular way if one examines DE and
DM sources by means of the usual FLRW models and their linear perturbations.             

\section{Interaction between the mixture components}

As we mentioned before, the two mixture components: DM (inhomogeneous
dust) plus DE (homogeneous fluid) are not separately conserved.
Considering (\ref{Theta}) and (\ref{CZ}), the energy balance for
the total energy--momentum tensor, \,$\dot e+(e+p)\,\Theta=0$, \,
can be written in terms of the DM and DE components as
\begin{equation} \left[\dot\rho+\rho\left(\,3\, H +\CZ\,\dot
T\right)\right]+\left[\dot\mu +(\mu+p)\left(\,3\, H +\CZ\,\dot
T\right) \right]= 0,\label{Ebal}\end{equation}
Since each term in square brackets in the left hand side of (\ref{Ebal})
corresponds to the energy balance of each mixture component alone, a
self--consistent form for describing the interaction between the latter is given by
\bse\label{Ebal2}\ba \dot\mu+(\mu+p)\left[\,3\, H
+\CZ\,\dot T\right] &=& -Q ,\\
\dot\rho+\rho\left[\,3\, H +\CZ\,\dot T\right] \ &=& Q, \ea\ese
where $Q=Q(t,r)$ is the interaction term. Since the physics
behind DM and DE remains so far unknown, we cannot  rule out the existence of such
interaction. Notice that once a given model has been
determined by specifying an ``equation of state'' $p=p(\mu)$ and a form for
$f=f(r)$, as we have done in the previous sections, this interaction term would also be
fully determined. In general, if the interaction
term in (\ref{Ebal2}) is a negative valued function, then DM transfers energy into
the DE and viceverza. Considering the free parameters given by (\ref{glaw}), 
(\ref{fform}) and (\ref{freepars}), we plot in figure 9 the interaction term $Q$ in
(\ref{Ebal2}), as a function of $\tau$ and $\log_{10} r$. Notice that $Q$ is
initially positive and remains so today (DE trasfers energy to DM at $\tau=\tau_0$) but
will change sign in a future time (DM trasfers energy to DE), tending to zero
asymptotically as $\tau\to\infty$. Thus, the time--asymptotic state is that of only
gravitational interaction between DE and DM ({\it i.e.} separate conservation of each
component).

However, the relevant question is not so much the explicit functional
form of $Q$, but its interpretation in terms of a self--consistent physical
theory that would be regulating the interaction between DM and DE. In fact, one of the
challenges of modern cosmology is to propose such a self--consistent theoretical model of
this interaction, while agreeing at the same time with the experimental and observational
data. In this context, the interaction between DM and DE has been considered, using
homogeneous FLRW cosmologies, in trying to understand the so--called ``coincidence
problem'', that is, the suspiciously coincidental fact the DE and DM energy densities are
of the same order of magnitude in our present cosmic era
\cite{amendola,coinc1,Q-int1,Q-int2,Q-int3,Q-int4,Q-int5}.    

If the homogeneous DE fluid corresponds to a quintessense scalar field,  $\phi=\phi(t)$,
with self-interaction potential $V(\phi)$, we have instead of (\ref{glaw}):
\begin{equation}\mu \ = \ \frac{\dot\phi^2}{2}+V(\phi),\qquad p \ = \
\frac{\dot\phi^2}{2}-V(\phi).\label{SFdef}\end{equation}
In this case, the interaction term in (\ref{Ebal2}) can be associated with a well
motivated non--minimal coupling to the DM component. 
Consider a scalar-tensor theory of gravity, where the matter
degrees of freedom and the scalar field are coupled in the action
through the scalar-tensor metric
$\chi(\phi)^{-1}g_{ab}$\cite{kaloper}:
\ba S_{\textrm{ST}} \ = \ \int d^4x
\sqrt{|g|}\{\frac{R}{2}-\frac{1}{2}(\nabla\phi)^2\nonumber\\+ \ \chi(\phi)^{-2}
{\cal L}_m(\nu,\nabla\nu,\chi^{-1}g_{ab})\}, \label{staction} \ea
where $\chi(\phi)^{-2}$ is the coupling function, ${\cal L}_m$ is
the matter Lagrangian and $\nu$ is the collective name for the
matter degrees of freedom. Equations (\ref{Ebal2}) become
\bse\label{Ebal3}\ba \ddot\phi+\dot\phi\, \left[\,3\, H +\CZ\,\dot
T\right]\ &=& -\frac{\dd V}{\dd \phi}
\ +\frac{\rho}{2\dot\phi}\frac{\dot\chi}{\chi},\label{Ebal4}\\
\dot\rho+\rho\left[\,3\, H +\CZ\,\dot T\right] \ &=& \
-\frac{\rho}{2}\;\frac{\dot\chi}{\chi}.\ea\ese
so that, the coupling function $\chi(\phi)$ and the interaction term $Q$ are
related by
\begin{equation}
Q=-\frac{\rho}{2}\;\frac{\dot\chi}{\chi}. \label{interactionterm}
\end{equation}
Therefore, once we determine a given model, so that $Q$ can be explicitly
computed, we can use (\ref{interactionterm}) to find the the coupling
function $\chi$ that allows us to relate the underlying interaction with
the theoretical framework associated with the action (\ref{staction}). 

For the models under consideration, we can integrate (in general) the constraint
(\ref{interactionterm}) with the help of (\ref{eq_eps}), (\ref{metric}), (\ref{eq_rho}),
(\ref{Theta}), (\ref{Ebal3}) and using 
$ \Theta=(d/dt)\,[\ln (Y^2\,Y')]$. This yields 
\ba \chi^{-1/2} &=& \ \xi(r)\,\rho\, Y^2\,Y'\nonumber\\
&=&
\frac{2c_0\,H_0\,r^2\,\xi(r)}{3\,a^3}\,\frac{d}{dt}\,\left[(2\,f\,F\,T+f+F)
\,R^3\right],\nonumber\\
\label{chi}\ea
where $\xi(r)$ is an arbitrary function that emerges as a constant of
integration.  Notice that the models require $\phi=\phi(t)$, so that the assumption
$\chi=\chi(\phi)$ implies $\chi=\chi(t)$. However, from (\ref{chi}), we have (in general)
$\chi=\chi(t,r)$, with the case $\chi=\chi(t)$ occurring for the following particular
cases, associated with very special forms of $f$ and $\xi$
\bse\label{particular1}\ba F &=& 0, \qquad f  \propto  r^{-3/2}, \quad \xi \propto
r^{-1/2},\nonumber\\
   &\Rightarrow& \chi^{-1/2} \propto \ H,\\\nonumber \\f+F&=&0, \qquad f  \propto 
r^{-3},
\quad \xi \propto r^4,\nonumber\\  &\Rightarrow& \chi^{-1/2} \propto \ \dot
T+3\,TH\ea\ese
or, if $F\ne 0 $ and $f+F\ne 0$, then $f$ and $\xi$ must be obtained from the
constraints:
\ba \frac{3}{2}\frac{f''}{f}+r\frac{f'{}^2}{f^2}+\frac{3f'}{f}+\frac{6}{r}
=0,\nonumber\\
\frac{\xi\,'}{\xi} = -\frac{6f+6 \,r
f'+r^2\,f''}{r(r\,f'+3f)}.\label{particular2}\ea\\
However, since (\ref{particular1}) and (\ref{particular2}) yield very
special forms of $f,\,\xi$ and of $\chi$, we prefer to apply (\ref{staction}) under
the most general assumption that the coupling function $\chi$ 
should be a function of $\phi$ and of position, {\textit{i.e.}} 
$\chi=\chi(\phi(t),r)$, as given by (\ref{chi}) for suitable forms of the free
functions $f$, $H$ and $T$ (thus, $\phi$ and $V(\phi)$), hence $\xi$ can be
considered a wholly arbitrary function. Considering the free parameters given by 
(\ref{glaw}), (\ref{glawpars}), (\ref{fform}) and (\ref{freepars}), and taking $\xi\propto
1/r^2$,  figure 10 illustrates qualitatively how the coupling function $\chi^{-1/2}$
given by (\ref{chi}) follows a power law time dependence, slightly modulated by the
change of local DM densities from the local overdensity region to the low density cosmic
background. Since the scalar field (\ref{SFdef}) that corresponds to the equation of
state  (\ref{glaw}) and the parameters (\ref{glawpars}) also exhibits a power law time
dependence~\cite{CJSF}, we do have approximately $\chi=\chi(\phi)$, as required by
(\ref{staction}). 

However, figure 10 is just a qualitative plot.  We should point out
that relating the interaction term, $Q$, to the formalism represented by (\ref{staction})
is strictly based on the formal similitude between the field equations derived from the
action (\ref{staction}), on the one hand, and equations (\ref{Ebal2}) and (\ref{SFdef}),
on the other. Also, the interaction between DM and DE in the context of (\ref{staction})
is severely constrained by experimental tests in the solar system~\cite{will}. 
A more detailed and carefull examination of the relation between $\chi$ and
$\phi$ that incorporates properly these points, as well as the application of
(\ref{staction}) to the models presented here, will be undertaken in future papers.

\section{Conclusion.}

We have presented a class of inhomogeneous cosmological models whose source is an
interacting  mixture of DM (dust) and a generic DE fluid. The relevance of the present
paper emerges when we realize that there are surprisingly few studies in which DE and DM
are the sources of inhomogeneous and anisotropic spacetimes (see \cite{Qinhom}), as
practically all study of the dynamical evolution of these components is carried on in the
context of homogeneous and isotropic FLRW cosmologies or linear perturbations on a FLRW
background. There are also very few papers that examine the possibility of
non--gravitational interaction between DE and DM. 

Once we assume or prescribe an ``equation of state'' ({\it i.e.} a relation
between pressure, $p$, and matter--energy density, $\mu$, of the DE fluid), we have a
specific DE model (quintessense scalar fields, tachyonic fluid, etc) and all the
time--dependent parameters can be determined by solving two coupled differential equations
reminissent of FLRW fluids. Since the spacial dependence of all quantities is governed by
the function $f=f(r)$, once the latter is selected the models become fully determined
(though, various important regularity conditions must be also satisfied: see section VI).
In order to work out this process we chose the simple ``gamma law'' equation of state:
$p=(\gamma-1)\mu$ (equation (\ref{glaw})), leading to analytic forms for all relevant
quantities, including the main observational parameters,
$\Omega_{\textrm{\tiny{DE}}},\,\Omega_{\textrm{\tiny{DM}}}$ and $q$. Our choice for a DE
fluid complying with (\ref{glaw}) is equivalent to a scalar field with exponential
potential, satisfying a scaling power law dependence on $t$~\cite{CJSF}. Although this is
a very idealized quintessense model, our aim has been to use it as a guideline that
illustrates how more sophisticated DE scenarios can be incorporated in future work
involving the models. 

As we have mentioned in section VI, the models homogeneize and isotropize asymptotically
in cosmic time for all fundamental observers and/or assumptions on the DE fluid, thus they
are well suited for studying the interaction between DE and DM in the context
of the evolution of large scale inhomogeneities (of the order of the scale of homogeneity
$\sim 100-300$ Mpc). By selecting appropriate boundary conditions, we can examine
inhomogeneities at various scales and/or asymptotic conditions (see figures 2 and 3). In
particular, we have explored the case of a local DM overdense region, whose scale can be
arbitrarily fixed and with an asymptotic behavior that accurately converges to a cosmic
background characterized by observational parameters that fit currently accepted
observational constraints: 
$0.6<\Omega_{\textrm{\tiny{DE}}}< 0.8,\,\,0.2<\Omega_{\textrm{\tiny{DM}}}< 0.4$ and
$q\approx -0.5$ (see figure 4). As illustrated by the various graphical examples that we
have presented, this interplay between a local overdensity, a cosmic background and a
transition region between them, shows in a spectacular manner  how inhomogeneity and
anisotropy lead to  interesting and important information that cannot be appretiated in
models based on FLRW metrics and/or linear perturbations. For example, as
revealed by figure 5, the effect of anisotropy emerges as a dependence of observational
parameters on local observation angles in ``off center postions'', an effect which is only
significant in the transition between the overdensity and the cosmic background. On the
other hand, as shown by figures 6-8, inhomogeneity allows for radically different ratios
between DM and DE in the overdensity and the cosmic background, so that DM dominates over
DE, locally, in the overdense region, as a contrast with DE dominating DM,
asymptotically, in the cosmic background (as expected). Also, while $q$ is negative in
the cosmic (DE dominated) background, thus denoting the expected ``repulsive''
accelerated  expansion at large scales, we have $q>0$ along smaller scales in the local
overdensity. For all parameters there is a smooth convergence between local and
asymptotic values in the transition region.

We have also examined the non--gravitational interaction between DE and DM. Ploting the
interaction term, $Q$, (figures 9) shows that energy flows from DE to DM
at the present cosmic era, with the flow reversing direction in the future and evolving
towards an asymptotic future state characterized by pure gravitational
interaction: $Q\to 0$. If we take the DE fluid to be a quintessense scalar field, the DE
vs DM interaction can be incorporated to the theoretical framework of an action like
(\ref{staction}), associated with a non--minimal coupling of scalar fields and DM. Since
DM is inhomogeneous while the scalar field is homogeneous, only for some particular forms
of spacial dependence ({\it i.e.} the function $f$)  we obtain a coupling function
expressible as $\chi=\chi(\phi)$. In general, we have to allow for the possibility that
$\chi=\chi(\phi(t),r)$, but as shown by figure 10, the spacial dependence of
$\chi$ under the assumption (\ref{glaw}) (scalar field with exponential potential)
mantains a power law dependence that is qualitatively very similar to that shown in the
homogeneous case. However, we have examined this interaction just in qualitative terms,
with the purpose of illustrating the methodology to follow in future applications.

As guidelines for future work, we have the application of the models to more
sophisticated and better motivated DE formalisms, perhaps in the context of the
``coincidence'' problem~\cite{amendola,coinc1,Q-int1,Q-int2,Q-int3,Q-int4,Q-int5}. An
interesting development would be the study of the case $K\ne 0$, in which hypersurfaces
of constant $t$ (rest frames) have nonzero curvature (see section II). This requires
solving the non--linear equation (\ref{ddotX}), which would certainly need numerical
techniques and perhaps assuming an homothetic symmetry that could reduce it to an ordinary
differential equation. The advantage would be the possibility of describing the effects
of DE on structure formation, since a DM overdense region with $K>0$ could collapse,
locally, as a DE dominated cosmic background with $K\leq 0$ expands. Finally, we have 
examined the DE/DM mixtures that can be constructed with the Szekeres--Szafron models of
``class 1'', as those of \cite{old2F2}, thus it still remains pending the study of ``class
2'' models like those of \cite{old2F1}.        

 \acknowledgements
RAS acknowledge finatial support from grant PAPIIT--DGAPA--IN117803.

\end{document}